\documentclass[a4paper,10pt,english, aps, prc, floatfix,  twocolumn,amsmath,amssymb,showpacs,notitlepage,nofootinbib]{revtex4-1}
\usepackage{graphicx}
\usepackage{mathrsfs}
\usepackage[sort&compress]{natbib}
\usepackage[unicode=true,pdfusetitle,
bookmarks=true,bookmarksnumbered=false,bookmarksopen=false,
breaklinks=false,pdfborder={0 0 1},colorlinks=false]{hyperref}
\hypersetup{colorlinks,citecolor=blue,linkcolor=blue}
\begin{document}

	\title{Cumulants of multiplicity distributions in most-central heavy ion collisions}

\author{Hao-jie Xu}

\email{haojiexu@pku.edu.cn}

\affiliation{Department of Physics and State Key Laboratory of Nuclear Physics
and Technology, Peking University, Beijing 100871, China}

\date{\today}

\begin{abstract} 
I investigate the volume corrections on cumulants of total charge distributions and net proton distributions.
The required volume information is generated by an optical Glauber model. I find
	that the corrected statistical expectations of multiplicity distributions mimic
	the negative binomial distributions at non-central collisions, and they tend to
	approach the Poisson ones at most-central collisions due to the "boundary effects,"
	which suppress the volume corrections. However, net proton distributions and reference
	multiplicity distributions are sensitive to the external volume fluctuations
	at most-central collisions, which imply that one has to consider the details of 
	volume distributions in event-by-event multiplicity fluctuation studies.

\end{abstract}


\maketitle 

\section{Introduction} 
In search of critical end point (CEP) in Quantum Chromodynamics (QCD) phase diagram~\cite{Aoki:2006we,Lacey:2014wqa,Gupta:2011wh,Stephanov:1999zu,Jeon:1999gr,Bazavov:2012vg,Gupta:2011wh},
event-by-event multiplicity fluctuations~\cite{Stephanov:1999zu,Jeon:1999gr,Bazavov:2012vg,Gupta:2011wh}
have been regarded as a very useful tool in relativistic heavy ion experiments~\cite{Adare:2015aqk,Tang:2014ama,Adamczyk:2014fia,Aggarwal:2010wy,Adamczyk:2013dal}.
Besides the fluctuation data and theoretical studies on critical fluctuations~\cite{Stephanov:2008qz,Chen:2014ufa,Chen:2015dra,Jiang:2015hri},
it is clear that a sufficient understanding of non-critical statistical fluctuations is also
important~\cite{Cleymans:2004iu,Begun:2004gs,Begun:2006uu,Garg:2013ata,Karsch:2010ck,Alba:2014eba,Fu:2013gga,Bhattacharyya:2015zka,BraunMunzinger:2011dn,Karsch:2015zna,Bzdak:2012an}.
As one of the non-critical corrections, the volume corrections on the cumulants of
multiplicity distributions have been discussed in previous studies~\cite{Jeon:2003gk,Gorenstein:2008th,Skokov:2012ds,Gorenstein:2011vq,Gorenstein:2013nea}.
Especially in Ref.~\cite{Skokov:2012ds}, Skokov and his collaborators have derived a general formalism for the
corrected cumulants. However, quantitative estimations of the 
volume corrections on data are obscure in previous statistical studies.

The effect of volume corrections on multiplicity distributions are caused by two reasons: (1) the distributions of volume
in heavy ion collisions (HIC), and (2) the multiplicity fluctuations of reference
particles used for centrality definitions. To quantitatively
estimate this effect in theoretical studies, therefore, the multiplicity fluctuations of
fluctuation measures and reference particles, as well as the volume distributions, need to be investigated.

In this work, I will continue my studies in Ref.~\cite{Xu:2016jaz}
(referred as [I]). In Ref.~[I], I have derived a general formalism for the multiplicity distributions
measured in experiments. With the corrected expression, I have calculated the volume
corrections under the Poisson approximations. I found that the statistical expectations
of multiplicity distributions mimic the negative binomial distributions (NBDs) at non-central
collisions. I also offer some reasonable explanations to the experimental data of (net) charge distributions.
It indicate the importance of volume corrections in heavy ion experiments.

In this work, I extend the study to the most-central collisions, i.e.,
the top few centrality percentage. Different from the platform-like distributions
in non-central collisions, the volume distributions at most-central collisions
 need more phenomenological considerations. This is due to the non-trivial
features of volume distributions at most-central collisions, which are reflected
in the rapid decreasing of the probability distributions of reference multiplicity when
the corresponding reference multiplicity is above an appropriate value. Such tail
distributions are caused by the finite size and energy of the colliding nuclei,
the fluctuating positions of the nucleons in the colliding nuclei, etc. In this work, I will use an optical Glauber
model to give more realistic descriptions of this information in relativistic
heavy ion collisions.

Meanwhile, the multiplicity distributions measured in most-central heavy ion experiments also
reveal some non-trivial features. Different from the large deviations at non-central
collisions, the scale variances of charged hadron distributions are close to
the trivial Poisson expectations at most-central collisions~\cite{Alt:2006jr,Adare:2008ns,Tang:2014ama}.
Moreover, the cumulant products of net proton distributions showed obvious
non-monotonic behavior at $0-5\%$ centrality percentage~\cite{Adamczyk:2014fia},
which was regarded as one of the most striking observables in searching for the CEP.
The $0-5\%$ centrality percentage is the transition range of the volume distributions,
and I will show in this work that the volume distributions at this centrality range
are very sensitive to the parameters of the phenomenological model. To study the corresponding
volume corrections, therefore, more realistic investigations of the volume distributions are required.
This, as far as I know, has not been addressed in previous statistical fluctuation studies.

The paper is organized as follows. In Sec.~\ref{sec:main}, I will give the main formalism
used in this work. The
analytical properties of charge fluctuations have also been discussed with a toy volume distribution.
In Sec.~{\ref{sec:results}} I show some numerical results on the cumulants of total charge
fluctuations, and discuss the effect of  Gaussian-type external volume fluctuations on net proton distributions.
I will give a summary in the final section.

\section{The model}
\label{sec:main}
In this section, I first review the main formalism of volume corrections given in [I] by using another expression,
and then I employ an Optical Glauber model to describe the required volume distributions in cumulant calculations.
At last, some analytical solutions are discussed by using a toy volume distribution.

\subsection{Volume corrections}
To bridge the gaps between experimental measurements and theoretical calculations, I have derived a general expression in  a statistical model [I]
for recent data on multiplicity distributions
at RHIC~~\cite{Aggarwal:2010wy,Adamczyk:2013dal,Adler:2007fj,Adare:2008ns,Adamczyk:2014fia,Tang:2014ama,Adare:2015aqk}.
The conditional probability distribution for the distribution of multiplicity $q$
in a given reference multiplicity bin $k$ reads
\begin{equation} 
	\mathscr{P}_{B|A}(q|k)=\frac{\int d\mathbf{\Omega}F(\mathbf{\Omega})P_{B}(q;\mathbf{\Omega})P_{A}(k;\mathbf{\Omega})}{\mathscr{P}_{A}(k)},
	\label{eq:generaleq} 
\end{equation}
with $\mathscr{P}_{A}(k)$ the  distribution of reference multiplicity
\begin{equation} 
	\mathscr{P}_{A}(k) = \int d\mathbf{\Omega}F(\mathbf{\Omega})P_{A}(k;\mathbf{\Omega}).\label{eq:genrefmultiplicity} 
\end{equation}
Here $q$ represents the multiplicity of
moment-analysis particles in sub-event $B$, and $k$ represents the multiplicity
of particles for the centrality definition in sub-event $A$. The latter $k$ is also called
reference multiplicity.
$P_{A}(k;\mathbf{\Omega})$ and $P_{B}(q;\mathbf{\Omega})$
stand for the corresponding  multiplicity distributions in a specific statistical ensemble
with a set of principal thermodynamic variables $\mathbf{\Omega}$.
I have assumed that the sub-event $A$ and  the sub-event $B$ are
independent of each other in each event (thermal system).

If only consider the distribution of system volume $V$, Eq.~(\ref{eq:generaleq}) and Eq.~(\ref{eq:genrefmultiplicity}) can be written as
\begin{eqnarray} 
	\mathscr{P}_{B|A}(q|k) &=&\frac{1}{\mathscr{P}_{A}(k)}\int dV F(V)P_{A}(k;V)P_{B}(q;V)\nonumber \\
	\mathscr{P}_{A}(k)  &=& \int dV F(V)P_{A}(k;V).
	\label{eq:distribution}
\end{eqnarray}
In experiments, total charged hadrons are usually chosen  as reference particles for
centrality definitions~\cite{Aggarwal:2010wy,Adamczyk:2013dal,Adler:2007fj,Adare:2008ns,Adamczyk:2014fia,Tang:2014ama,Adare:2015aqk}. 
As shown in [I], the multiplicity fluctuations of total charges can be described
by the Poisson distributions in a fixed volume. So  $P_{A}(k;V)$ can be described by a Poisson distribution,
then the volume $V$ ( volume distribution $F(V)$) can be substituted by the Poisson parameters $\lambda$ ($f(\lambda)$).
Consequently,
\begin{eqnarray} 
	\mathscr{P}_{B|A}(q|k) &=& \mathscr{N}(k)\int d\lambda f(\lambda)\frac{\lambda^{k}e^{-\lambda}}{k!} P_{B}(q;\lambda), \nonumber \\
	\mathscr{P}_{A}(k)  &=& \int d\lambda f(\lambda)\frac{\lambda^{k}e^{-\lambda}}{k!},
	\label{eq:distribution}
\end{eqnarray}
where  $\mathscr{N}(k)=1/\mathscr{P}_{A}(k)$ is normalization factor.

Similar to Ref.~\cite{Skokov:2012ds}, the first four cumulants of $\mathscr{P}_{B|A}(q|k)$ can  be written as
\begin{eqnarray}
	\label{eq9}
	\frac{c_{1}}{\langle\lambda\rangle}&=&\kappa_{1}, \nonumber\\
	\frac{c_{2}}{\langle\lambda\rangle}&=& \kappa _2 + \kappa _1^2 v_2, \nonumber\\
	\frac{c_{3}}{\langle\lambda\rangle}&=& \kappa _3+3 \kappa _2 \kappa _1 v_2+ \kappa _1^3 v_3, \label{eq:cumulants}\\
	\frac{c_{4}}{\langle\lambda\rangle}&=& \kappa _4 +(4 \kappa _3 \kappa _1+3 \kappa _2^2 )v_2+
6 \kappa _2 \kappa _1^2 v_3 + \kappa _1^4 v_4 \nonumber,
\end{eqnarray}
Here $\kappa_{1},\kappa_{2},\kappa_{3}$ and $\kappa_{4}$ are the first four reduced cumulants of $P_{B}(q;\lambda)$ 
\begin{eqnarray}
	\kappa_{1} &=& \frac{\sum q P_{B}(q;\lambda)}{\lambda}\equiv \frac{\bar{q}}{\lambda}, \nonumber \\
	\kappa_{2} &=& \frac{\sum (\Delta q)^{2} P_{B}(q;\lambda)}{\lambda}, \nonumber \\
	\kappa_{3} &=& \frac{\sum (\Delta q)^{3} P_{B}(q;\lambda)}{\lambda},   \\
	\kappa_{4} &=& \frac{ \sum(\Delta q)^{4} P_{B}(q;\lambda) - 3 (\sum (\Delta q)^{2}P_{B}(q;\lambda))^{2}}{\lambda}, \nonumber
\end{eqnarray}
with $\Delta q\equiv q - \bar{q} $,
\begin{eqnarray}
	v_{2} &=& \frac{\langle(\Delta\lambda)^{2}\rangle}{\langle\lambda\rangle}, \nonumber\\
	v_{3} &=& \frac{\langle(\Delta\lambda)^{3}\rangle}{\langle\lambda\rangle}, \\
	v_{4} &=& \frac{\langle(\Delta\lambda)^{4}\rangle -3\langle(\Delta\lambda)^{2}\rangle^{2}}{\langle\lambda\rangle}, \nonumber 
\end{eqnarray}
with $\Delta\lambda=\lambda-\langle\lambda\rangle$, $\langle(...)\rangle\equiv\mathscr{N}(k)\int d\lambda
f(\lambda)\frac{\lambda^{k}e^{-\lambda}}{k!}(...)$ and
\begin{eqnarray}
	\langle\lambda^{m}\rangle &=& \mathscr{N}(k)\int d\lambda f(\lambda)\frac{\lambda^{k}e^{-\lambda}}{k!}\lambda^{m} \nonumber \\
																			&=& \frac{(k+m)!}{k!} \frac{\mathscr{P}_{A}(k+m)}{\mathscr{P}_{A}(k)}.  
	\label{eq:lambdam}
\end{eqnarray}

For the charge fluctuations in this work, I assume $P_{B}(q;\lambda)$ is a Poisson distribution with parameter $\mu=\bar{q}=b\lambda$. The reduced cumulants read
\begin{equation}
	\kappa_1 = \kappa_2 = \kappa_3 = \kappa_4 = b.
\end{equation}
For the net conserved charge fluctuations, I assume $P_{B}(q;\lambda)$ is a Skellam distribution with parameters $\mu_{+}=b_{+}\lambda$ for the positive conserved charges, and $\mu_{-}=b_{-}\lambda$ for the negative conserved charges. The reduced cumulants read
\begin{equation}
	\kappa_1 = \kappa_3 = b_{+}-b_{-}\equiv \Delta b;\ \ \  \kappa_2 = \kappa_4 = b_{+} + b_{-}  \equiv b_\pm.
\end{equation}
The values of $b$, $b_{+}$ and $b_{-}$  are related to the multiplicity ratios of different particle
species with different acceptance cuts, which can be described by  statistical model if the kinematic cuts on particles of interest can
be well simulated. For example, in the net proton case, there is a simple relation between $b_{-}$ and $b_{+}$
in classical statistical model with grand canonical ensemble, $b_{-}/b_{+}=\exp{(-2\mu_{B}/T)}$,
where $\mu_{B}$ and $T$ are baryon chemical potential and temperature.

In non-central heavy ion collisions, one has $\mathscr{P}_{A}(k+m)/\mathscr{P}_{A}(k)\simeq 1$ for small $m$ and
\begin{equation}
	\langle\lambda\rangle=k+1,\ \  v_{2}=1,\ \  v_3=2,\ \  v_{4}=6.
\end{equation}
Then the first four cumulants of total charge distributions can be written as
\begin{eqnarray} 
	\frac{c_{1}}{k+1} & = & b \equiv M/(k+1) ,\label{eq:sigma2total} \nonumber\\ 
	\frac{c_{2}}{k+1} & = & b+b^{2},\label{eq:sigma2total}\nonumber\\ 
	\frac{c_{3}}{k+1} & = & b + 3b^{2} + 2b^{3},\label{eq:charge}\\ 
	\frac{c_{4}}{k+1} & = & b + 7b^{2} + 12 b^{3} + 6 b^{4}. \nonumber
\end{eqnarray} 
And the first four cumulants of net conserved charge distributions can be written as
\begin{eqnarray} 
	\frac{c_{1}^{N}}{k+1} & = & \Delta b \equiv \frac{M_{+}}{k+1}-\frac{M_{-}}{k+1},\nonumber\\ 
	\frac{c_{2}^{N}}{k+1} & = & b_\pm + \Delta b^{2},\nonumber\\
	\frac{c_{3}^{N}}{k+1} & = & \Delta b + 3b_\pm\Delta b + 2\Delta b^{3},\label{eq:netcharge}\\
	\frac{c_{4}^{N}}{k+1} & = & b_\pm+4\Delta b^{2}+3b_{\pm}^{2}+12b_\pm\Delta b^{2}+6 \Delta b^{4}. \nonumber 
\end{eqnarray} 
Eq.~(\ref{eq:charge}) and Eq.~(\ref{eq:netcharge}) are the appropriate solutions
obtained in [I] at non-central collisions.
To obtain the exact values of cumulants in a given reference multiplicity 
bin $k_0$, one only need the information of mean multiplicity of moment analysis particles,
similar to the Poisson/Skellam expectations. For the (net) charge distributions
reported by the STAR collaboration, the volume corrections are important because the reduced
cumulants $b$ ($b_{+},b_{-}$) are of the order of $O(1)$. Therefore the volume corrections
play a crucial role to describe the negative binomial multiplicity distributions of (net)
charges~[I].

Some remarks are in order here. In the above discussions, I have assumed that $b$, $b_{+}$ and $b_{-}$
are independent of centrality (volume). This is because in typical statistical
models, the Poisson parameters $\mu$ ($\mu_{+}$,$\mu_{-}$), $\lambda$ are both proportional
to volume. However, this assumption can be contaminated by various effects,
such as: 
(1) the un-platform distribution of multiplicity as function of
pseudo-rapidity and (2) the uncertainties from experiment inefficiencies, etc.
To reduce the uncertainties from centrality-dependent $b$ ($b_{+}$, $b_{-}$),
therefore,
I suggest the following centrality definition approach. Firstly, each event is divided into
two sub-events with different (pseudo-)rapidity cuts: $y_{\mathrm{min}}<y<y_{\mathrm{max}}$
and $-y_{\mathrm{max}}<y<-y_{\mathrm{min}}$. Here a rapidity gap $2y_{\mathrm{min}}$ is
employed to reduce the auto-correlation effect~\footnote{In this work, the auto-correlation means the correlation
of sub-event A and B in one event. It can be generated by collective flow, hadronization, jet,
resonance decays, etc. This is different from the one discussed in the present work, i.e.
the correlation from event-by-event analysis.} of these two sub-events in each event. These kinematic cuts will
significant reduce the uncertainties from source (1) in symmetric nuclear collisions.  Secondly,
randomly choose one of the sub-events for centrality
definition, and leave the other one for moment-analysis in each event. This approach will largely
suppress the contributions from source (2). Finally, with the role reversal of these
two sub-events, all the events can be used to calculate the corresponding cumulants twice.
In this sense this approach redoubles the statistics, and it is extremely useful due to
the statistic hungry properties of the cumulant calculations.
Note that this approach can not reduce the  non-linear contributions from fluctuations of intensive variables, e.g. $\mu_{B}$ and $T$.

\subsection{Volume distributions}
As mentioned above, the volume distributions have been substituted by
$f(\lambda)$, the distributions of  Poisson parameters $\lambda$.
The reason that $\mathscr{P}_{A}(k+m)/\mathscr{P}_{A}(k)\simeq 1$
is used in [I] at non-central collisions is that the volume distributions are
platform-like distributions at the corresponding centrality range.
To give the cumulant calculations at most-central collisions, however,
the details of $f(\lambda)$ need to be investigated.

In this work an optical Glauber model~\cite{Kharzeev:1996yx,Kharzeev:2000ph,Kolb:2000sd,Abelev:2008ab,Miller:2007ri}
is used to simulate the distribution of $f(\lambda)$ (see Appendix~\ref{sec:glauber} for the details)
\begin{equation}
	f(\lambda) = \mathscr{R} \int P(n;\zeta)[1 - P_{0}(\zeta)]2\pi \zeta d\zeta 
	\label{eq:glauber}
\end{equation}
where $\mathscr{R}$ is a normalization factor and $\zeta$ is the impact parameter.
The correlation function can be written as~\cite{Kharzeev:2000ph}
\begin{equation}
	P(\lambda;\zeta) = \frac{1}{\sqrt{2\pi a\lambda}}\exp\left[-\frac{(\lambda-hn(\zeta))^2}{2a\lambda}\right]
\end{equation}
which stands for  Gaussian-type volume fluctuation~\footnote{The external volume fluctuation is very different
from the so-called "volume fluctuation" in previous Ref.~\cite{Skokov:2012ds}. The so-called "volume fluctuation" is actually
the volume correction discussed in previous subsection.
While the external volume fluctuation is the fluctuating term of the
Glauber volume distribution. In hydrodynamics, it can be generated by fluctuating initial conditions, fluctuating freeze-out hyper-surfaces, etc.
The volume corrections still exist in the absence of external volume fluctuations.
}.
The parameter $a$ is the strength of fluctuation. In the absence of fluctuation $a=0$, the Gaussian distribution become to a Delta function. Here
\begin{equation}
	n(\zeta) = \left[\frac{(1-x)}{2}n_{\mathrm{part}}(\zeta) + xn_{\mathrm{coll}}(\zeta)\right]/n(0) 
\end{equation}
and $n_{\mathrm{part}}(\zeta)$ is the number of participant nucleons,
$n_{\mathrm{coll}}(\zeta)$ is the number of binary nucleon-nucleon collisions.
The fitting parameter $h$ is determined by the size and energy of
the colliding nuclei. The fraction parameter
is chosen as $x=0.12$ in this work. $ P_{0}(\zeta)$ is the probability
of no interaction among the nuclei~\cite{Kharzeev:2000ph}
\begin{equation}
	P_{0}(\zeta) = \left[ 1 - \frac{n_{\mathrm{coll}}(\zeta)}{N_{A}N_{B}}\right]^{N_{A}N_{B}}
\end{equation}
with mass number of the collisional nuclei $N_{A}$ and $N_{B}$.

\subsection{Analytical solutions}
\label{sec:analysis}
Before presenting the numerical results of cumulant calculations in the next section,
I now discuss some analytical properties of the multiplicity distributions by using a toy volume
distribution.

Assume that the distribution $f(\lambda)$  is a platform distribution in the
range $\lambda\in[0,h]$, i.e.,
\begin{equation}
	f(\lambda)=\begin{cases}
		1/h & \lambda\leq h,\\
		0 & \lambda>h,
	\end{cases}
\end{equation}
where $h$ correspond to the upper boundary of the system volume.

From above assumptions and the Poisson approximation of $P_{B}(q;\mu)$ with parameter $\mu=b\lambda$, Eq. (\ref{eq:distribution})
can be rewritten as 
\begin{eqnarray}
	\mathscr{P}_{B|A}(q|k) &=&\frac{b^{q}}{q!}\frac{\int_{0}^{h}d\lambda\lambda^{k+q}e^{-(1+b)\lambda}}{\int_{0}^{h}d\lambda\lambda^{k}e^{-\lambda}} \nonumber \\
																						  &=& \frac{1}{q!}\left(\frac{b}{1+b}\right)^{q}\left(1-\frac{b}{1+b}\right)^{k+1} \nonumber \\
																  & & \times\frac{\gamma(k+q+1,h(1+b))}{\gamma(k+1,h)},
\end{eqnarray}
where $\gamma(s,x)$ is the lower incomplete Gamma function~\footnote{The author thanks T. S. Biro for a discussion of this point.}.
\begin{enumerate}
	\item At non-central collisions, $k+q\ll h$, 
\begin{equation}
	\mathscr{P}_{B|A}(q|k)\simeq \mathrm{NBD}(q;r,p)\equiv \frac{(r+q-1)!}{(r-1)!q!}p^{q}(1-p)^{r},
	\label{eq:NBDexp}
\end{equation}
which mimics a standard negative binomial distribution with parameters $r=k+1$
and $p=b/(1+b)$. These results are consistent
with the approximate solutions given in [I] (see also Eq.~(\ref{eq:charge}) in this work).

 \item  At most-central collisions, $k\gg h$,
\begin{eqnarray}
	\mathscr{P}_{B|A}(q|k) & \simeq & \frac{e^{-bh}(bh)^{q}}{q!}\frac{k+1}{k+q+1},
\end{eqnarray}
where I have used
\begin{equation}
	\gamma(s,x) = x^{s}(s-1)!e^{-x}\sum_{m=0}^{\infty}\frac{x^{m}}{(s+m)!}
\end{equation}
and the leading order approximation
\begin{equation}
	\sum_{m=0}^{\infty}\frac{x^{m}}{(s+m)!} \simeq \frac{1}{s!}
\end{equation}
when $x/s\ll1$.
For the distribution with $q\ll k$,
\begin{equation}
	\mathscr{P}_{B|A}(q|k)\sim\frac{e^{-bh}(bh)^{q}}{q!},
\end{equation}
which mimics a Poisson distribution with parameter $\tilde{\mu}=bh$.
\end{enumerate}

Therefore, as the reference multiplicity $k$ increases,
the distribution $M(k)$ will be saturated at
high reference-multiplicity range because the parameter $\tilde{\mu}$
is independent of $k$. Moreover, the variances will change the value from the NBD
expectations to the Poisson predictions. These features have
been also observed in experiments~\cite{Alt:2006jr,Adare:2008ns,Tang:2014ama},
which indicate that the volume corrections are weak in most-central
heavy ion collisions. The suppression of
volume corrections at most-central collisions has been found
in previous work~\cite{Skokov:2012ds} but with so-called
symmetric volume fluctuations. The different contributions of volume
corrections on cumulants of multiplicity distributions in non-central 
and most-central collisions are due to the tail distributions of volume
at its upper boundary and I call this the "boundary effects" in this work.

\section{Results and discussions}
\label{sec:results}
In this section I present the results of the corrected  cumulants of total charge distributions
and net proton distributions~\footnote{The relations between the net proton distributions
and its corresponding net conserved charge  distributions, i.e. net baryon distributions,
have been discussed in Ref.\cite{Kitazawa:2011wh,Kitazawa:2012at}.}.

The free parameters in the model are: the multiplicity ratios $b$ ($b_{+}$ and $b_{-}$),
the Glauber parameters $h$ and $a$, and some other parameters in the optical Glauber model.
To determine these values at given collision energy, the information of reference multiplicity
distribution $\mathscr{P}_{A}(k)$ and the means multiplicity distributions of the fluctuation measures $\mathscr{M}(k)$ are required.
Unfortunately, both of them are not available in current experiments. Instead of realistic baseline predictions,
the main purpose of this work is investigating the sensitivity of higher order cumulants of multiplicity fluctuations on
the collision geometry at most-central collisions.
Therefore, as the first attempt, the fitting parameter of the optical Glauber model in this work is set to $h=400$ and the
strength parameter of external Gaussian fluctuation is set to $a=0.01$ as default choice unless stated explicitly.
For the net proton distribution, the Poisson parameter for proton distribution is set to $\mu_{+} = b_{+}\lambda = 0.04\lambda$,
and the one for anti-protons distribution is set to $\mu_{-}= b_{-}\lambda = 0.01\lambda$.

\begin{figure*}
	\begin{center} 
		\includegraphics[scale=0.4]{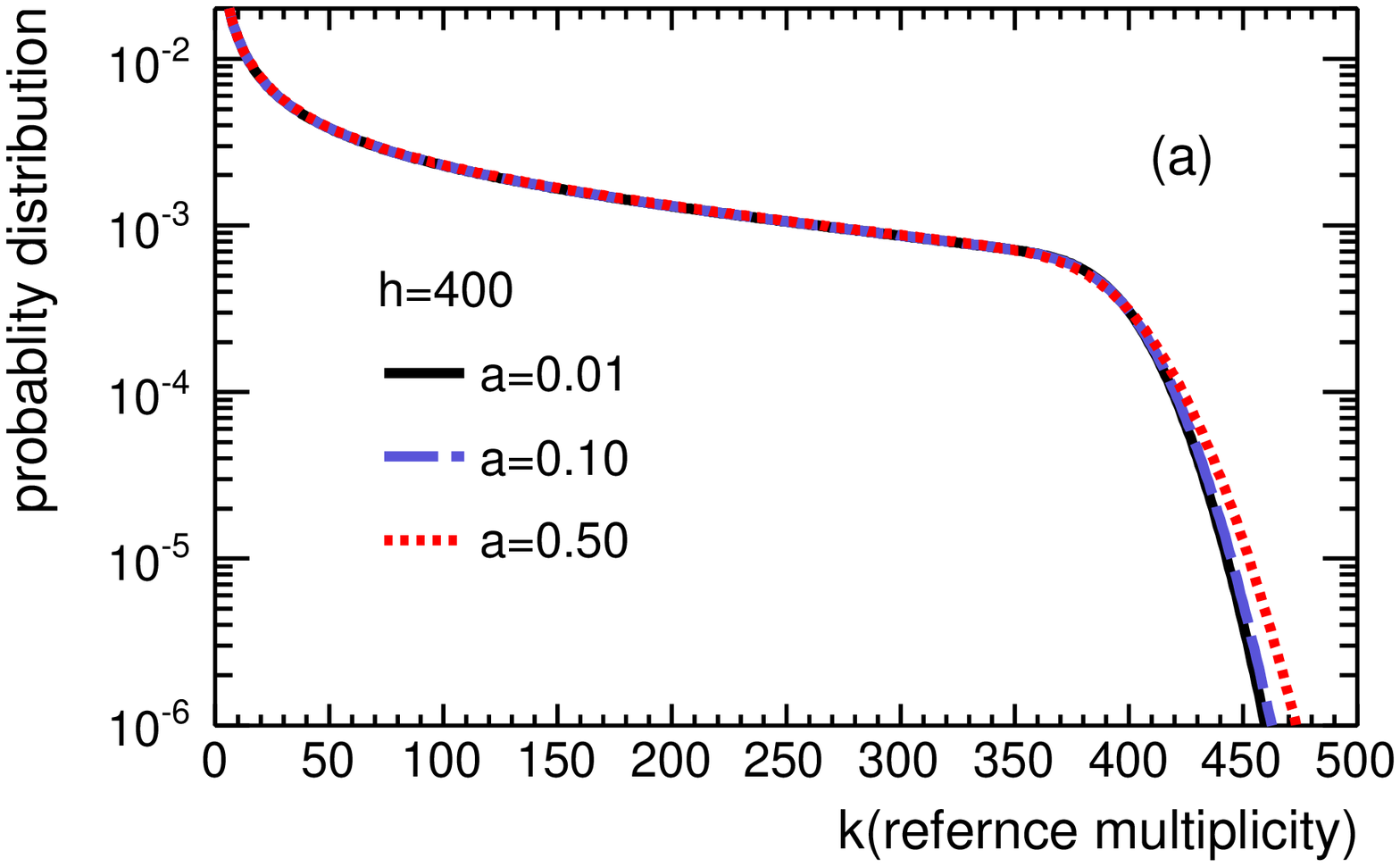} \includegraphics[scale=0.4]{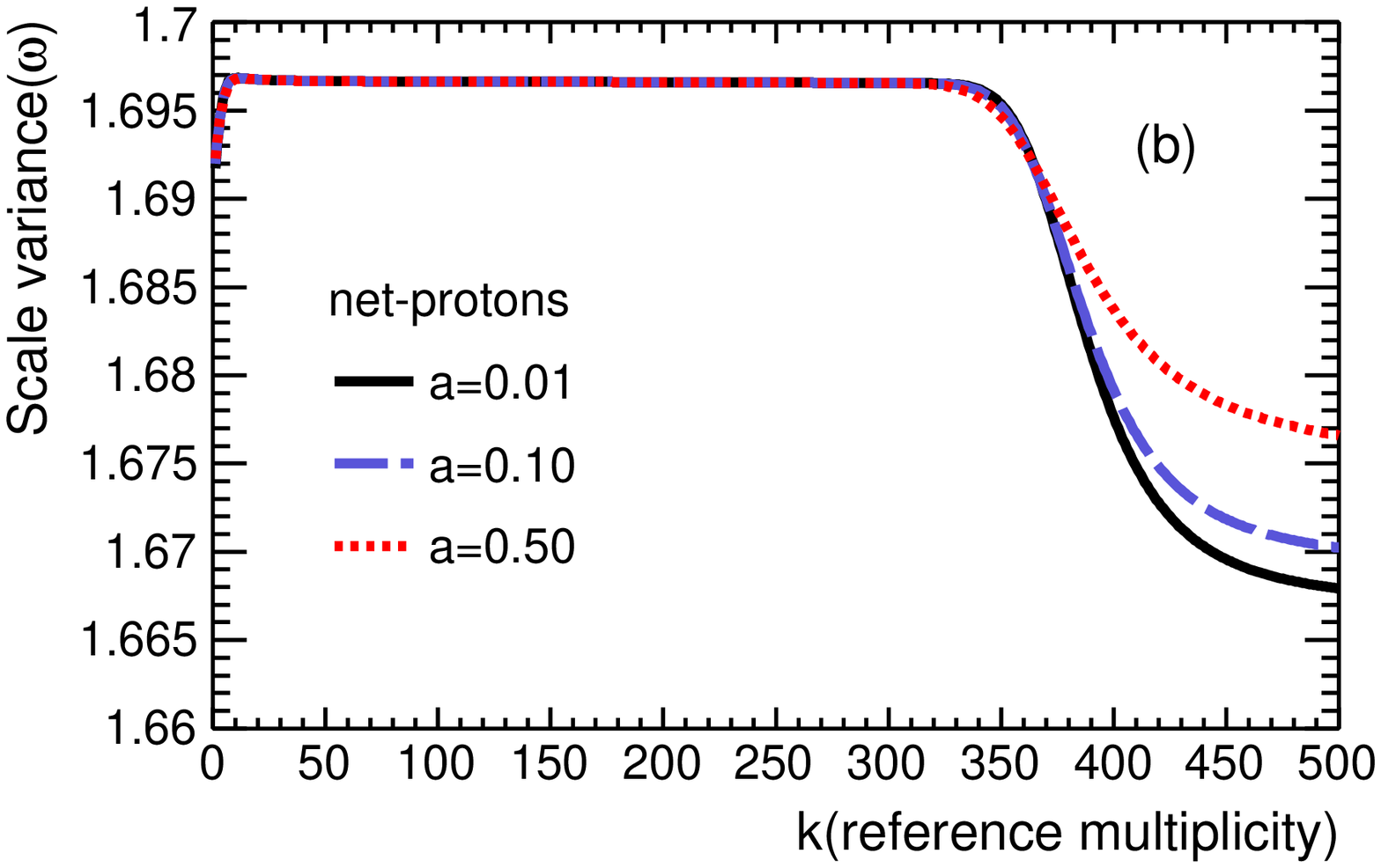} \\
		\includegraphics[scale=0.4]{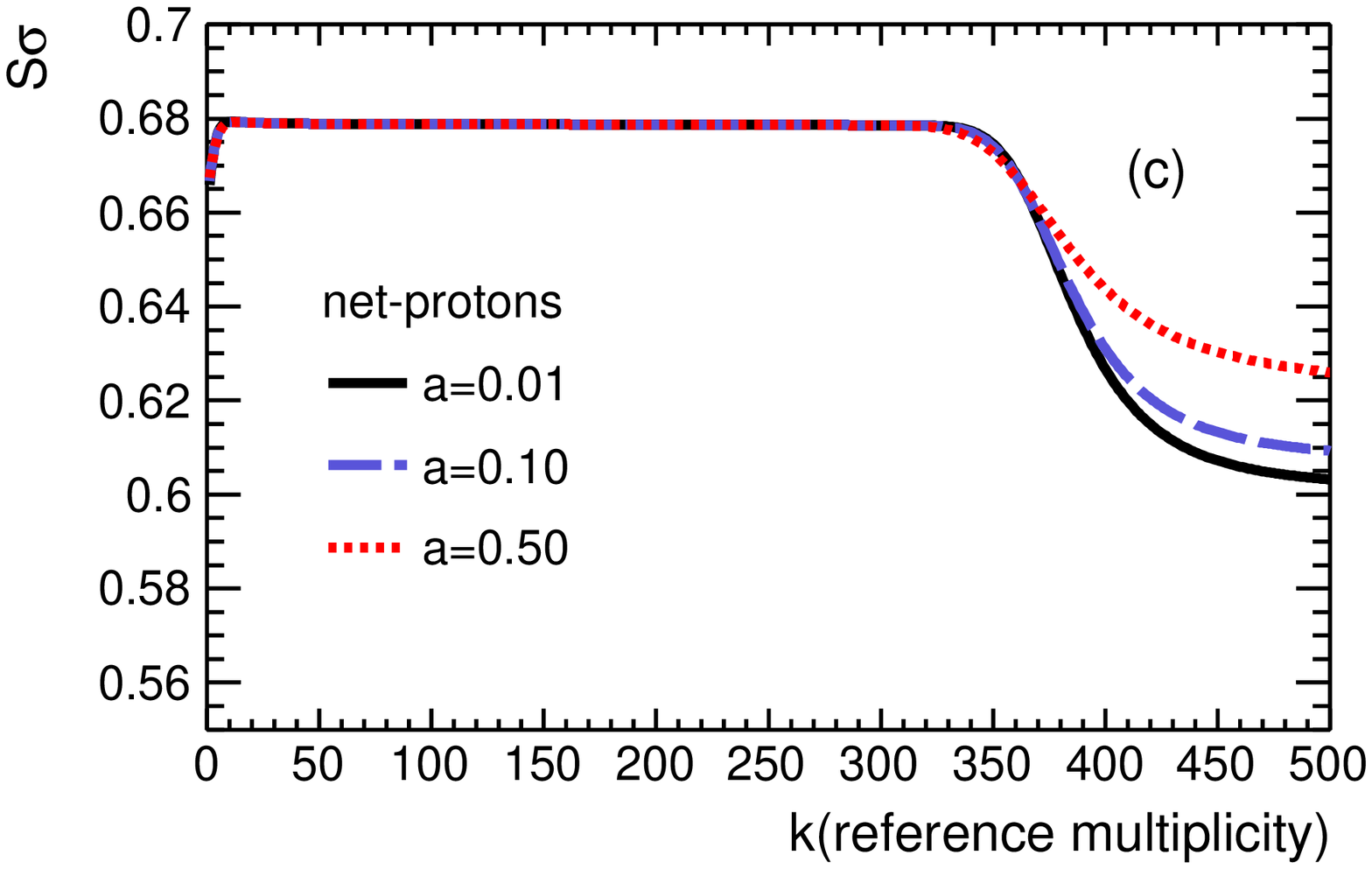} \includegraphics[scale=0.4]{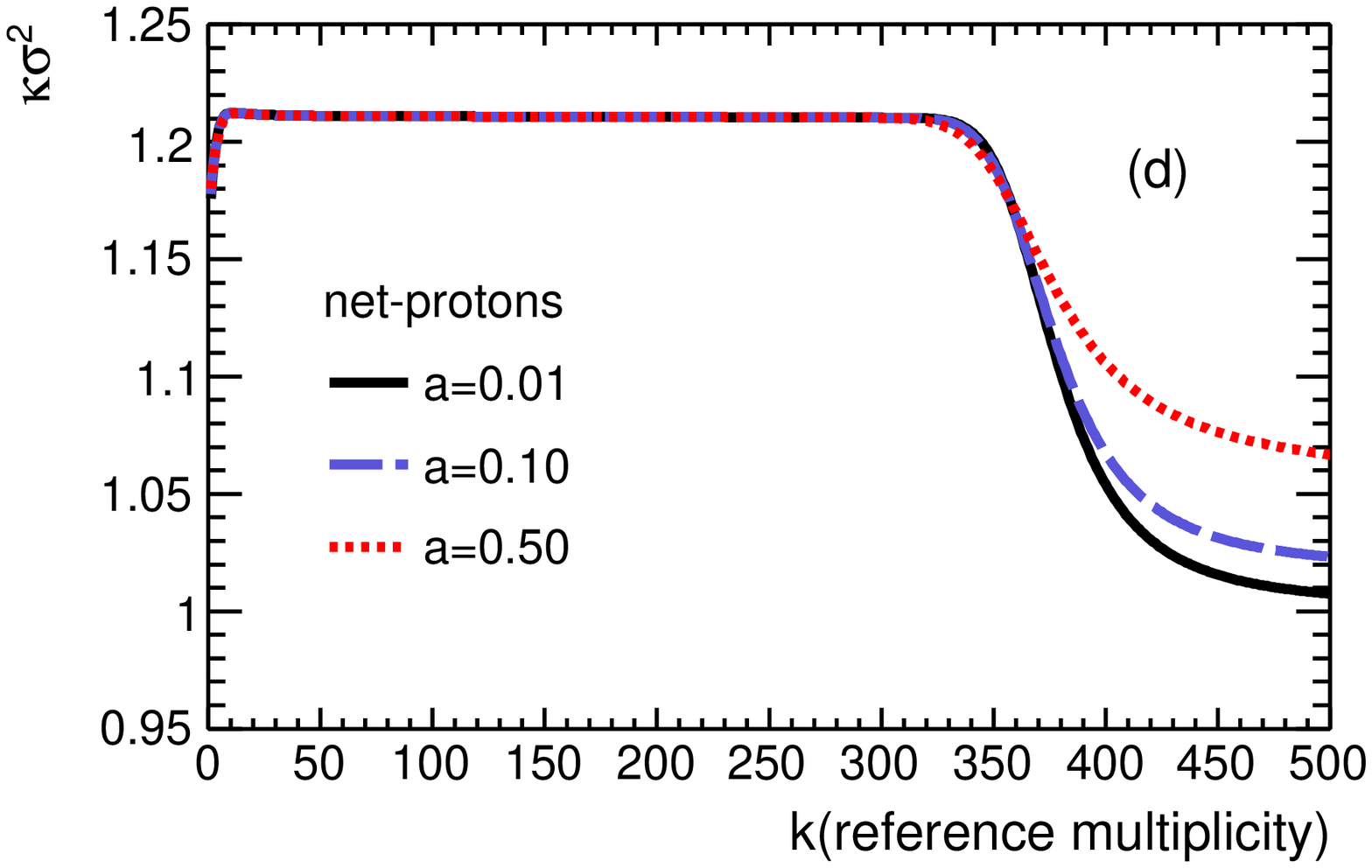} 
	\end{center} 
	\caption{(Color online) (a) Distributions of reference multiplicity and cumulant products (b) $\omega=c_2/c_1$, (c) $S\sigma=c_3/c_2$, (d) $\kappa\sigma^2=c_4/c_2$ of net proton distributions with various strength of Gaussian-type external volume fluctuations $a=0.01$ (black-solid lines), $a=0.10$ (blue-dashed lines) and $a=0.50$ (red-dotted lines).	\label{fig:comparison}} 
\end{figure*}

Fig.~\ref{fig:siglevalue}(a) shows the distribution of Poisson
parameter $f(\lambda)$ and the corresponding  distribution of reference
multiplicity $\mathscr{P}_{A}(k)$ (see also the black-solid curve in Fig.~\ref{fig:comparison}(a)).
The contributions of event-by-event Poisson fluctuations on
$\mathscr{P}_{A}(k)$ are obvious at high reference multiplicity range, i.e. most-central collisions.

\subsection{Total charges}

\begin{figure*}
	\begin{center} 
		\includegraphics[scale=0.4]{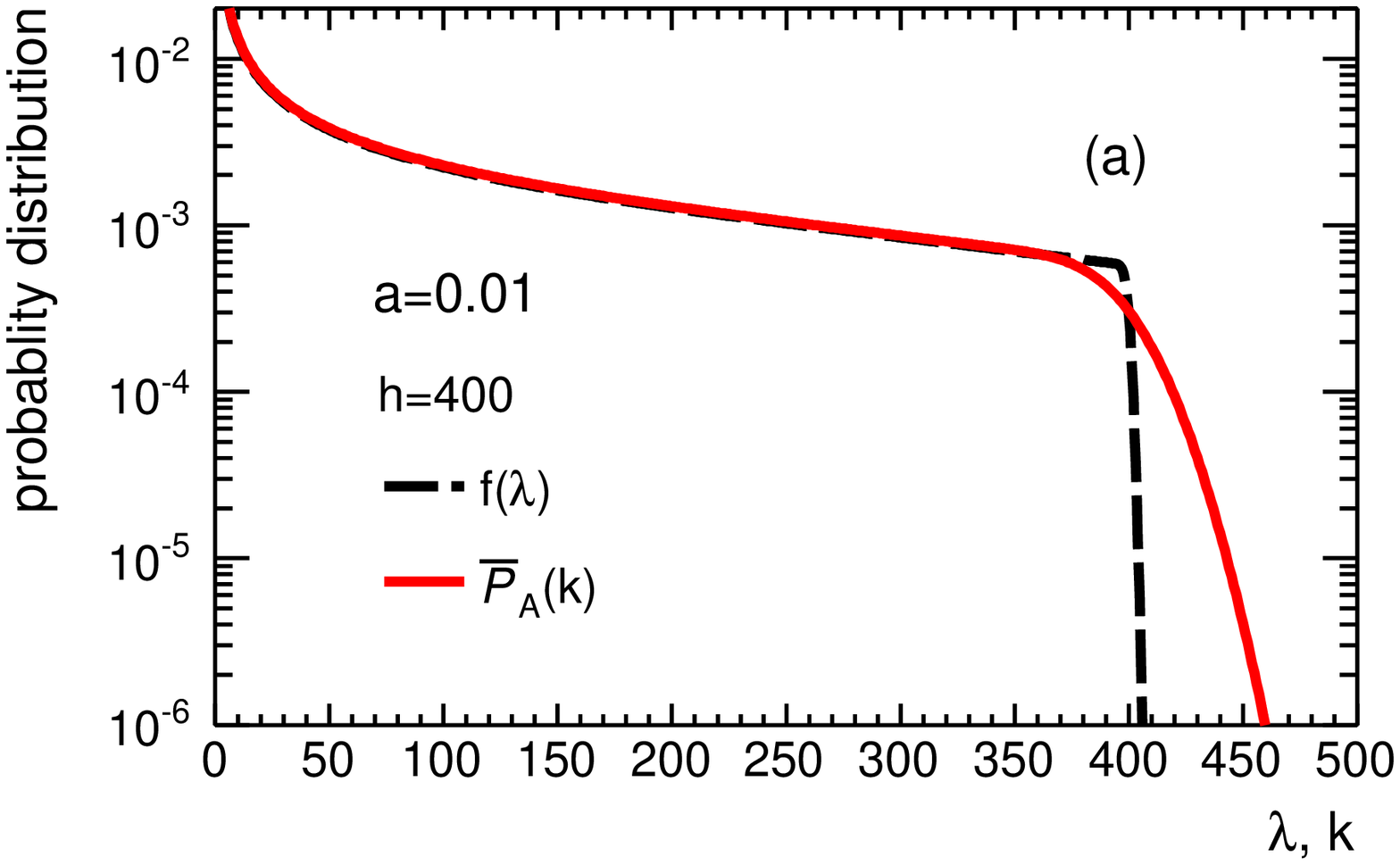}
		\includegraphics[scale=0.4]{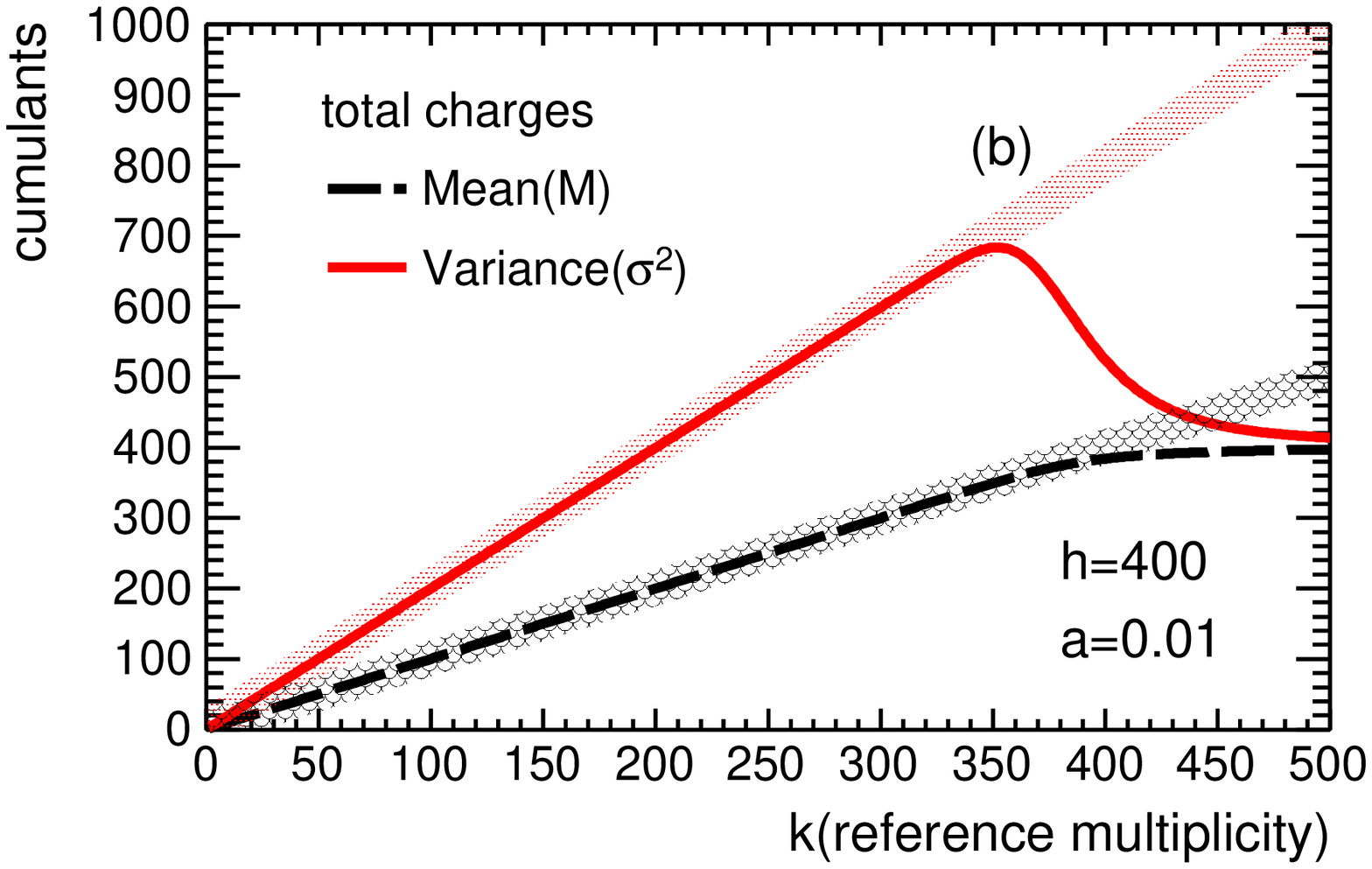}\\
		\includegraphics[scale=0.4]{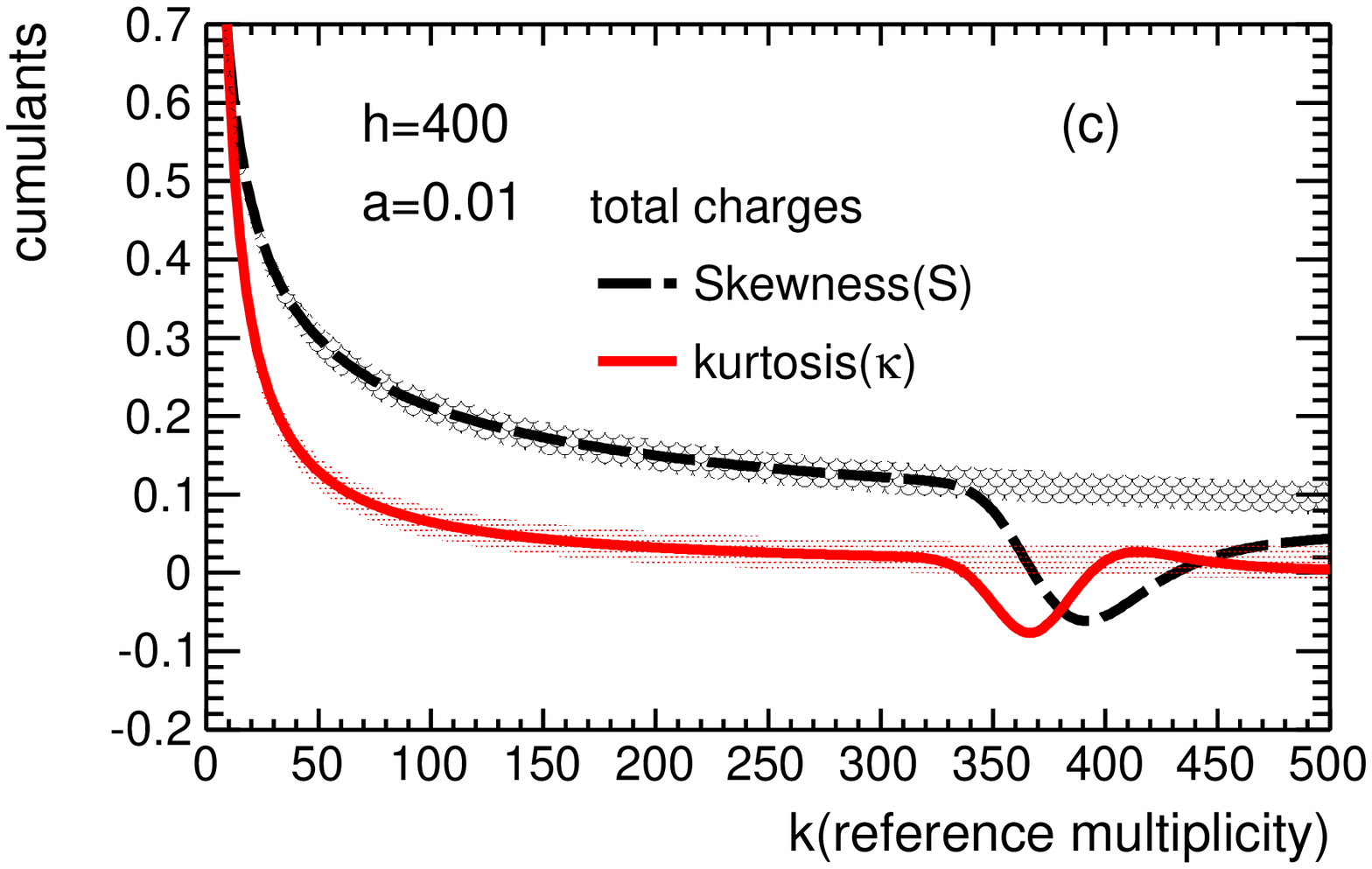}
		\includegraphics[scale=0.4]{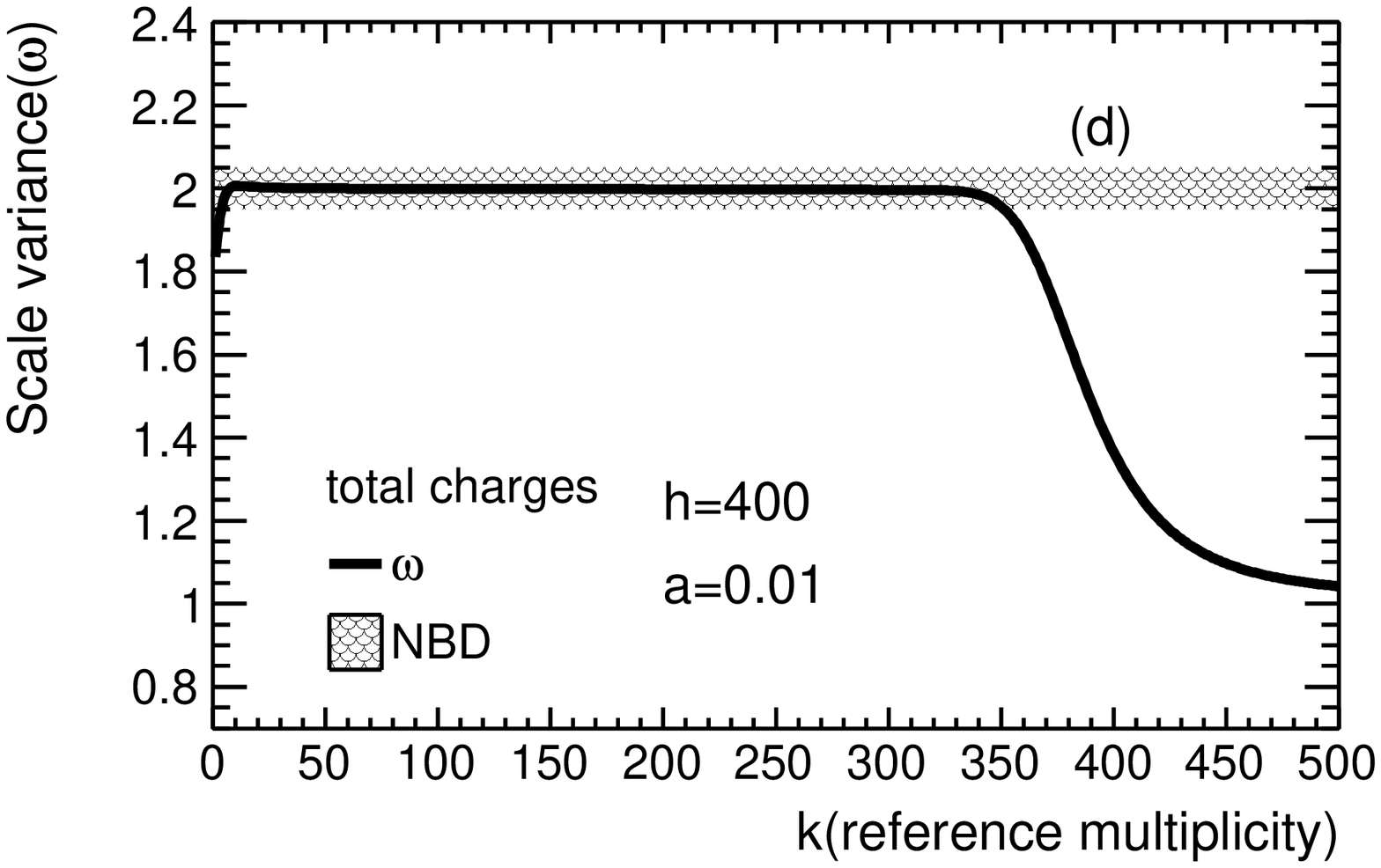}
	\end{center} 
	\caption{(Color online) (a) Probability distribution of Poisson parameter $\lambda$ (black-dashed line) and
		reference multiplicity (red-solid line). (b). Mean value ($M$), variance ($\sigma^{2}$), and (c) skewness($S$)
		, kurtosis ($\kappa$) of total charge fluctuations. (d) The scale variance ($\omega=\sigma^{2}/M$) of total charge
	distributions. The notation $\bar{P_{A}}(k)$ in (a) stand for $\mathscr{P}_{A}(k)$ in the text. The bands in (b), (c), (d) represent the corresponding NBD expectations from Eq.~(\ref{eq:NBDexp}).
	\label{fig:siglevalue}} 
\end{figure*}

In the total charges
case, one has $b=1$ if the centralities are
also defined by the multiplicity of total charges with the recommended centrality definition approach.
The numerical results for cumulants of total charge distributions are shown in Fig.~\ref{fig:siglevalue}(b)
and Fig.~\ref{fig:siglevalue}(c). The corresponding bands represent the NBD expectations carried out by Eq.~(\ref{eq:NBDexp}).

The numerical results can be well described by the NBD expectations in a wide centrality range,
which are consistent with my previous conclusions~[I]. At most-central collisions, due to the "boundary effects",
the mean values of total charge distributions tend to saturate in the high
reference multiplicity range. Meanwhile, the variances of total charge distributions
tend to approach its mean value, which make the distributions mimic the Poisson distributions
at second order moment level.

To further investigate this transition, Fig~\ref{fig:siglevalue}(d) shows
the scale variance of total charge fluctuations $\omega=\sigma^2/M$.
It is quite clear that from peripheral to central collisions,
the scale variances of total charge fluctuations vary from being NBD expectations
($\omega=1+M/(k+1)\simeq2$) to Poisson expectations ($\omega=1$).
The deviations of scale variances from the Poisson expectations at
high reference multiplicity range (ultra-central collisions)
are due to the external Gaussian fluctuation on collision geometry,
see Eq.~(\ref{eq:glauber}).

It is interesting to find that the centrality dependence of skewness
and kurtosis have non-monotonic behavior in high reference multiplicity range, and
the minimum values of skewness and  kurtosis can be negative.
Although the volume correction effects are suppressed by the "boundary effects",
the non-trivial behavior of high order cumulants of total charge distributions imply that
realistic descriptions of the volume distributions (e.g. collision geometry, etc.) are
important in studying the transition of multiplicity
distribution from non-central to most-central heavy ion collisions.

\subsection{Net protons}

Now I focus on fluctuations of net protons.
I calculate the cumulant products  $\omega\equiv c_{2}/c_{1}$, $S\sigma\equiv c_{3}/c_{2}$
and $\kappa\sigma^{2}\equiv c_{4}/c_{2}$ of net proton distribution
as shown in Fig.~\ref{fig:comparison}(b), Fig.~\ref{fig:comparison}(c) and Fig.~\ref{fig:comparison}(d) (black-solid lines).

Except for the most-central collision range, the numerical results are consistent with
the approximate solutions given in [I], i.e.,
\begin{eqnarray}
	\omega &=& \beta(1-\alpha) + \frac{1+\alpha}{1-\alpha} = 1.697, \\
	S\sigma & = & 2\beta(1-\alpha)+\frac{\beta(1-\alpha^{2})+1-\alpha}{\beta(1-\alpha)^{2}+1+\alpha} = 0.679,\label{eq:Ssigma}\\
	\kappa\sigma^{2} & = & 6\beta(\gamma-\frac{2\alpha}{\gamma})+1 = 1.211,
\end{eqnarray}
where $\alpha=M_{-}/M_{+}=b_{-}/b_{+}$, $\beta=M_{+}/(k+1)=b_{+}$ and $\gamma=\beta(1-\alpha)^{2}+1+\alpha$.
The approximate solutions of these cumulant products are independent of the centralities.
In general, the cumulant products
can been used to extract the chemical freeze-out parameters from hadron resonance gas model~\cite{Bazavov:2012vg,Alba:2014eba}.
However, I note that the effect of volume corrections need to be taken into account for these extractions.

Although I can not determine the sources and magnitude of the Gaussian-type external volume fluctuations,
it is still interesting to study the effect of such external
fluctuations on the measured cumulants. I therefore calculate the cumulant products  of net proton distributions with various strength
parameter $a=0.01$(default), $a=0.10$, and $a=0.50$, respectively.
Note that for some specific calculations at given collision energy,
these parameters in Glauber model can be constrained by the data
of $\mathscr{P}_{A}(k)$ if it becomes available.

Fig.~\ref{fig:comparison}(a) shows that the distribution of reference multiplicity has a wider tail for a larger strength parameter $a$.
The effect of different external volume fluctuations can be only distinguished at most-central collision range, see Fig.~{\ref{fig:comparison}}(b-d).
The reason is since that for the cumulants of net proton distributions calculated
from Eq.~(\ref{eq:cumulants}), the contributions from  distribution of volume was related to the ratio of
$\mathscr{P}_{A}(k+m)/\mathscr{P}_{A}(k)$, see Eq.~(\ref{eq:lambdam}). The differences of
this ratio in different scenarios are obvious only at most-central collisions, see Fig.~\ref{fig:comparison}(a).
It is worth noting that, the magnitude
of the external Gaussian-type fluctuation's strength can be reflected not only in the second order cumulant ($\omega$) of net proton distribution,
but also in the more sensitive higher order ones, i.e, $S\sigma$ and $\kappa\sigma^{2}$.
The results imply that,
besides the non-Gaussian critical fluctuations suggested in Ref.~\cite{Stephanov:2008qz},
the non-monotonic behavior
of higher order cumulants beyond variances at high reference multiplicity range could be also generated by volume corrections
with different Gaussian-type volume fluctuations.

In the future, I will consider several other effects in multiplicity fluctuation studies, e.g, 
quantum effect, resonance decays, experimental acceptance, the correlation between  different moment-analysis particles, as well as the
correlation between moment-analysis particles and reference particles, etc. As I have discussed in [I],
these corrections are especially important in studying of net proton fluctuations. Therefore, instead of an exploratory study given in this work,
more elaborate studies are desired to pin down the exact statistical predictions of measured cumulants in heavy ion collisions.

\section{Conclusions} 

I have extended my previous work of corrected cumulants of (net conserved) charge distributions 
to most-central heavy ion collisions. The required volume distributions are simulated by an optical Glauber model. Under the Poisson approximations,
I calculated the corrected cumulants of total charge
(net proton) distributions from a general formalism according to the data.
To reduce the uncertainties between experimental measurements and theoretical calculations, as well as redouble
the statistics, I also suggested a special approach for centrality definition.

I found that the statistical expectations of multiplicity distribution mimic the NBD at non-central collisions,
but tend to approach the Poisson one at most-central collisions. This transition is because of the "boundary effects" that were
caused by the upper boundary of the system volume in HIC,
which significant suppress the volume corrections.
I have further investigated the effect of external Gaussian-type volume fluctuations
on the high order cumulants of net proton distributions.

The results indicate that the non-critical volume corrections on high order
cumulants of multiplicity distributions  become weak at most-central collisions.
However,
the sensitivity of net proton distribution and reference
multiplicity distributions on the external volume fluctuations
implies that the details of volume distribution
in relativistic heavy ion collisions need to be
considered carefully in event-by-event multiplicity fluctuation studies.

\section*{Acknowledgments}  
The author thanks Shi Pu for a careful reading of the manuscript and useful suggestions.
The author also thanks V. Koch, K. Redlich and N. Xu for fruitful discussions at the CPOD 2016 conferences.
This work is supported by China Postdoctoral Science Foundation under Grant No.~2015M580908.

\appendix

\section{Optical Glauber model}
\label{sec:glauber}
The density distribution of the colliding nuclei in Glauber model~\cite{Kharzeev:1996yx,Kharzeev:2000ph,Kolb:2000sd,Abelev:2008ab,Miller:2007ri}
is given by  Woods-Saxon profiles,
\begin{equation}
	\rho_{N}(r) = \frac{\rho_0}{\exp\left[(r-R_{N})/\xi\right]+1},
\end{equation}
with the nuclear radius $R_{N} = (1.12N^{1/3} - 0.86N^{-1/3})$ fm, the normal nuclear density $\rho_{0}=0.1699 \mathrm{fm}^{-3}$ and the
surface diffuseness $\xi=0.54$ fm. Here $N$ is mass number of the nuclei. The nuclear thickness function is obtained
from the optical path-length through the nucleus along the beam direction
\begin{equation}
	T_{N} = \int_{-\infty}^{+\infty}dz\rho_{N}(x,y,z).
\end{equation}

The number density of binary collisions with impact parameter $\zeta$ reads
\begin{equation}
	n_{\mathrm{bc}}(x,y;\zeta) =\sigma_{in} T_{N_{A}}(x+\zeta/2,y)T_{N_{B}}(x-\zeta/2,y),
\end{equation}
where $\sigma_{\mathrm{in}}$ is the total inelastic cross section. Then the total number of binary collisions is
\begin{equation}
	n_{\mathrm{coll}}(\zeta) = \int dxdy n_{\mathrm{bc}}(x,y;\zeta).
\end{equation}

The number density of wounded collisions (participants) with impact parameter $\zeta$ is
\begin{eqnarray}
	n_{\mathrm{wn}}(x,y;\zeta)&=&\left[1-\left.(1-\frac{\sigma_{in}T_{N_{B}}(x-\frac{\zeta}{2})}{N_{B}}\right.)^{N_{B}}\right] \nonumber \\
																							& &\times T_{A}(x+\frac{\zeta}{2},y)+ (N_{A} \leftrightarrow N_{B}).
\end{eqnarray}
where $N_{A}=N_{B}=197$ for Au+Au collisions. Then the total number of wounded collisions (participants) is
\begin{equation}
	n_{\mathrm{part}}(\zeta) = \int dxdy n_{\mathrm{wn}}(x,y;\zeta).
\end{equation}

\bibliography{ref}

\end{document}